\documentclass[12pt]{article}
\usepackage[margin=1in]{geometry}
\usepackage{float}

\usepackage{fullpage}

\usepackage{amssymb, amsthm, amsmath}
\usepackage{doublespace}
\usepackage{bm, url}
\usepackage{textcomp}
\usepackage{graphicx}
\usepackage[authoryear]{natbib}
\usepackage{bm}
\usepackage{verbatim}
\usepackage{lineno}
\usepackage{times}
\usepackage{caption}
\usepackage{subcaption}
\usepackage{epstopdf}
\usepackage{float}
\usepackage{bbm}
\usepackage{mathrsfs}

\newcommand{\btheta}{ \mbox{\boldmath $\theta$}}

\newcommand{\balpha}{ \mbox{\boldmath $\alpha$}}

\newcommand{\bSigma}{ \mbox{\boldmath $\Sigma$}}

\newcommand{\bff}{ \mbox{\bf f}}

\newcommand{\bX}{ \mbox{\bf X}}

\newcommand{\bH}{ \mbox{\bf H}}

\newcommand{\iid}{\stackrel{iid}{\sim}}

\newcommand{\calX}{{\cal X}}
\newcommand{\calZ}{{\cal Z}}

\newcommand{\calS}{{\cal S}}

\newcommand{\argmax}{{\mathop{\rm arg\, max}}}

\newcommand{\beq}{ \begin{equation}}
\newcommand{\eeq}{ \end{equation}}
\newcommand{\beqn}{ \begin{eqnarray}}
\newcommand{\eeqn}{ \end{eqnarray}}

\begin{document}

\begin{singlespace}
\title{Bayesian Nonparametric Policy Search with Application to Periodontal Recall Intervals}

\author{\small{Qian Guan$^{1,}$, Brian J. Reich$^{1}$, Eric B. Laber$^{1}$ and Dipankar Bandyopadhyay}$^{2}$\\ \\
{\small $^{1}$Department of Statistics, North Carolina State University, Raleigh, North Carolina}\\
{\small $^{2}$Department of Biostatistics, Virginia Commonwealth University, Richmond, Virginia}
}

\date{}
\maketitle

\end{singlespace}

\bigskip
\begin{singlespace}
\begin{abstract}
	
Tooth loss from periodontal disease is a major public health burden in the United States.  Standard clinical practice is to recommend a dental visit every six months; however, this practice is not evidence-based, and poor dental outcomes and increasing dental insurance premiums indicate room for improvement. We consider a tailored approach that recommends recall time based on patient characteristics and medical history to minimize disease progression without increasing resource expenditures.  We formalize this method as a dynamic treatment regime which comprises a sequence of decisions, one per stage of intervention, that follow a decision rule which maps current patient information to a recommendation for their next visit time. The dynamics of periodontal health, visit frequency, and patient compliance are complex, yet the estimated optimal regime must be interpretable to domain experts if it is to be integrated into clinical practice.  We combine non-parametric Bayesian dynamics modeling with policy-search algorithms to estimate the optimal dynamic treatment regime within an interpretable class of regimes. Both simulation experiments and application to a rich database of electronic dental records from the HealthPartners HMO shows that our proposed method leads to better dental health without increasing the average recommended recall time relative to competing methods.
\vspace{12pt}
	
\end{abstract}
\end{singlespace}

\newpage

\section{Introduction}\label{s:intro}

Periodontal disease (PD) contributes to eventual tooth loss and remains a major health burden. The ultimate goal of professional periodontal maintenance plans \citep{glossaryperio} and personal oral care is PD prevention and maintaining teeth in a state of comfort and function. The total dental healthcare spending in the United States in 2013 was a staggering US\$ 91.8 billion \citep{oralcost}, and is continually increasing \citep{cdc2011}.  Hence, there is an urgent need to reduce cost without diminishing the quality of care.  In the context of dental care, updating periodontal recall recommendations to reflect the individual needs of each patient holds the potential to both improve oral health and reduce cost. The length of periodontal recall intervals has been a topic of research and debate for decades \citep{lovdal1961combined, axelsson1991prevention, fardal2004tooth,
mettes2005insufficient, riley2013recall} with recommendations for recall intervals ranging from two weeks \citep{nyman1975effect} to eighteen months \citep{rosen1999effect}. The current standard of care, which was advocated as early as 1879 by the American Academy of Dental Science  \citep{teich2013risk}, is a recall interval of six months for all patients regardless of individual demographics, oral health, family history, or other risk factors.  Existing clinical guidelines recommend that the recall intervals should depend on individual patient characteristics (NCCAC, 2004; Patel et al., 2010; see also Giannobile et al., 2013) but offer little concrete guidance on how to map individual patient characteristics to a recall interval.

The potential effect of altering such intervals on oral health had remained the subject of international debate for almost 3 decades \citep{mettes2005insufficient, riley2013recall}. Infrequent dental visits might impair the ability to diagnose PD at an early stage, present fewer opportunities for providing oral-care education, and block the opportunities for effective treatments \citep{davenport2003effectiveness}. However, unnecessary visits and treatments waste the resources and increase the cost. Therefore, the recall interval should be tailored to individual needs. Patients at high risk may benefit from more frequent visits  while less frequent visits might be adequate for subjects without certain risk factors of PD \citep{giannobile2013patient}. This provides evidence that a personalized (or precision) medicine approach \citep{kornman2012personalized,zanardi2012future} might improve resource allocation for preventive dentistry.


In this paper, we consider adaptive recall intervals that recommend a recall time for each patient at each visit depending on their personal characteristics, including disease history. We formalize a personalized recall interval policy as a function that maps current patient information to a recommended recall interval, which is an example of a dynamic treatment regime, or DTR \citep[][]{murphy2003optimal, robins2004optimal, chakraborty2013statistical, schulte2014q, kosorok2015adaptive}. A DTR is defined as a sequence of decision rules, one per stage of intervention, to make treatment decisions based on the patient’s evolving status. Each decision rule takes the individual's information up to that time point as the input, and outputs a recommended treatment at that stage. The optimal DTR is defined as the regime that optimizes the mean long-term outcome. The problem we are addressing is to specify a regime that uses a patient's up-to-date information to tailor recall interval recommendations in such a way that maximizes long-term population-level benefits, and it is thus a DTR problem. DTRs have been applied across a wide range of application domains to estimate data-driven intervention policies \citep[][]{van2007causal, robins2008estimation, shortreed2012estimating, laber2014dynamic, almirall2014introduction, wu2015will}; however, estimation of an optimal recall intervention policy presents several challenges that make existing estimation methods unsuitable without modification. These challenges include: (i) cost constraints on recall frequency across the entire population; (ii) non-compliance and sparse irregularly spaced clinic visits; (iii) a bounded response with a large point mass on the response of the previous time point due to clinicians carrying forward previous measurements rather than  retaking them; and (iv) the requirement that the estimated policy be clinically interpretable, despite complex disease dynamics. Existing methods for cost-constrained DTRs include cost-constrained IQ-learning \citep[][]{linn2015constrained} which only applies for two decision points; cost-sensitive DTRs \citep[][]{luedtke2016optimal} which constrain the proportion of individuals who can receive treatment; set-valued DTRs \citep[][]{laber2014set, lizotte2016multi} which allow for multivariate outcomes, e.g., cost and efficacy, but do not permit constrained estimation.  Functional and longitudinal methods for DTRs can accommodate sparse and irregularly-spaced observation times \citep[][]{ciarleglio2015treatment,lu2015comparing, laberFnl}, however, these methods are not designed for application with many, possibly outcome-driven, follow-up times, nor can they deal with non-compliance.

Policy-search is a common method for estimation of a DTR, and is particularly well-suited to constrained problems \citep[][]{chakraborty2013statistical,wang2018learning,laber2018identifying}. Policy-search methods postulate a model for the marginal mean outcome under each policy within a pre-specified class of policies and choose the maximizer as the estimated optimal policy \citep[][]{robins2008estimation, orellana2010dynamic, zhang2012robust, zhang2012estimate, zhao2012estimating, zhang2013robust, zhao2015new, kosorok2015adaptive, guan2016discussion, zhou2017augmented}. An advantage of policy-search methods is that models for the underlying disease progression can be decoupled from the class of policies, thereby allowing for complex disease models with parsimonious, interpretable, or cost-constrained estimated optimal policies \citep[][]{zhang2015using, laber2015tree, lakkaraju2016learning}. However, existing methods for policy-search are difficult to implement for complex data structures, like the one we consider here.

We use a Bayesian nonparametric (BNP) disease dynamics model and g-computation \citep{robins1986new} to construct an estimator of the marginal mean outcome and cost under any policy within a pre-specified class, and then use stochastic optimization to approximate the maximizer of the mean outcome under a constraint on cost. The marginal mean outcome is cumulative, and accounts for disease progression and delayed effects of treatment. We estimate this marginal mean outcome using g-computation, which accounts for these effects and other time-varying causal confounding. The proposed dynamics model is sufficiently flexible to accommodate non-compliance, sparse and irregularly-spaced visits; however, our class of policies is based on a clinically interpretable risk score. BNP methods have recently been used in the context of estimating optimal treatment regimes \citep{arjas2010optimal, xu2016bayesian, murray2017robust}, but they did not consider regimes that adapt to the evolving health status of each individual patient, or cost constraints.


The motivation for establishing this (recall) recommendation engine comes from analyzing an observational database in a dental practice-based setting, collected by the HealthPartners\textregistered (HP) Institute at Minneapolis, Minnesota. In Section \ref{s:data}, we review the HP data. In Section  \ref{s:model}, we formalize the recall estimation problem using a decision theoretic framework. In Section \ref{s:BNP}, we present a BNP formulation of the disease dynamics and in Section \ref{s:policy}, we combine this model with a stochastic optimization algorithm to construct an estimator of the optimal intervention policy subject to constraints on cost. In Section \ref{s:sim}, we evaluate the finite sample performance of the proposed methods using a suite of simulation experiments. We analyze the motivating HP dataset and summarize the fitted policy in Section \ref{s:HP}. Finally, we conclude with a brief discussion in Section \ref{s:con}.


\section{HealthPartners Data}\label{s:data}

The motivating longitudinal HP dataset were collected from routine dental practice in the Minneapolis area. We include only adult subjects with at least two visits, giving 24,731 subjects with as many as 8 years of irregular longtitudinal follow-up, with an average of 8.6 visits. For each subject, we use the data from the first visit until the last visit to fit the model proposed in Section \ref{s:model}, and so the follow-up window varies by subject. During each visit, periodontal pocket depth (PPD) is recorded at six pre-specified sites per tooth (excluding the third molars) giving 168 measurements for a full mouth without any missing tooth. In concordance with the proposed standards from the joint EU/USA Periodontal Working Group \citep{holtfreter2015standards}, we use the proportion of diseased/affected tooth sites (with PPD $>$ 3mm, or missing tooth) per mouth, henceforth PMU, as our response to measure the extent (severity) of PD. Note, when the tooth is missing, we assume the missing is due to PD and we classify all sites associated with the missing tooth are diseased tooth sites. So each missing tooth contributes to 6 diseased tooth sites in the calculation. Demographic information and medical history are also collected, including age (ranging from 19 to 97 years, with mean 55 years), gender (49\% male, 51\% female), race (85\% white, 15\% non white), diabetes status (8\% with diabetes, 92\% without diabetes), smoking status (9\% current tobacco user, 91\% not current user), and insurance information (80\% with commercial insurance, 20\% without commercial insurance).

\begin{figure}
\caption{\small{The proportion of sites with unhealthy PPD (pocket depth exceeding 3mm) or missing tooth, denoted by PMU, over time, for 10 randomly chosen subjects.}}\label{f:subject}
\centering
\includegraphics[height=0.5\textwidth]{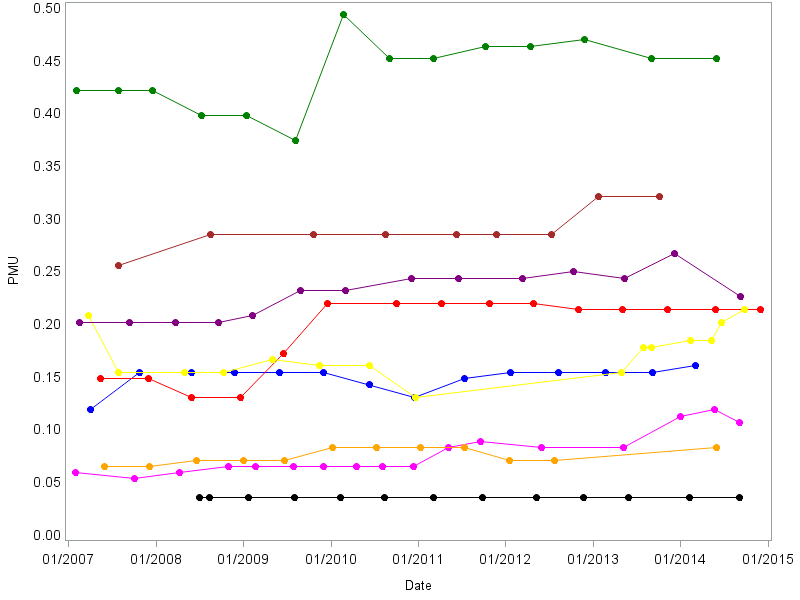}
\end{figure}

\begin{figure}
\centering
\caption{\small{Density plots of actual time between visits ($\delta$ months) for each level of recommended recall interval ($A$ months).}}\label{f:density}
\includegraphics[height=0.5\textwidth]{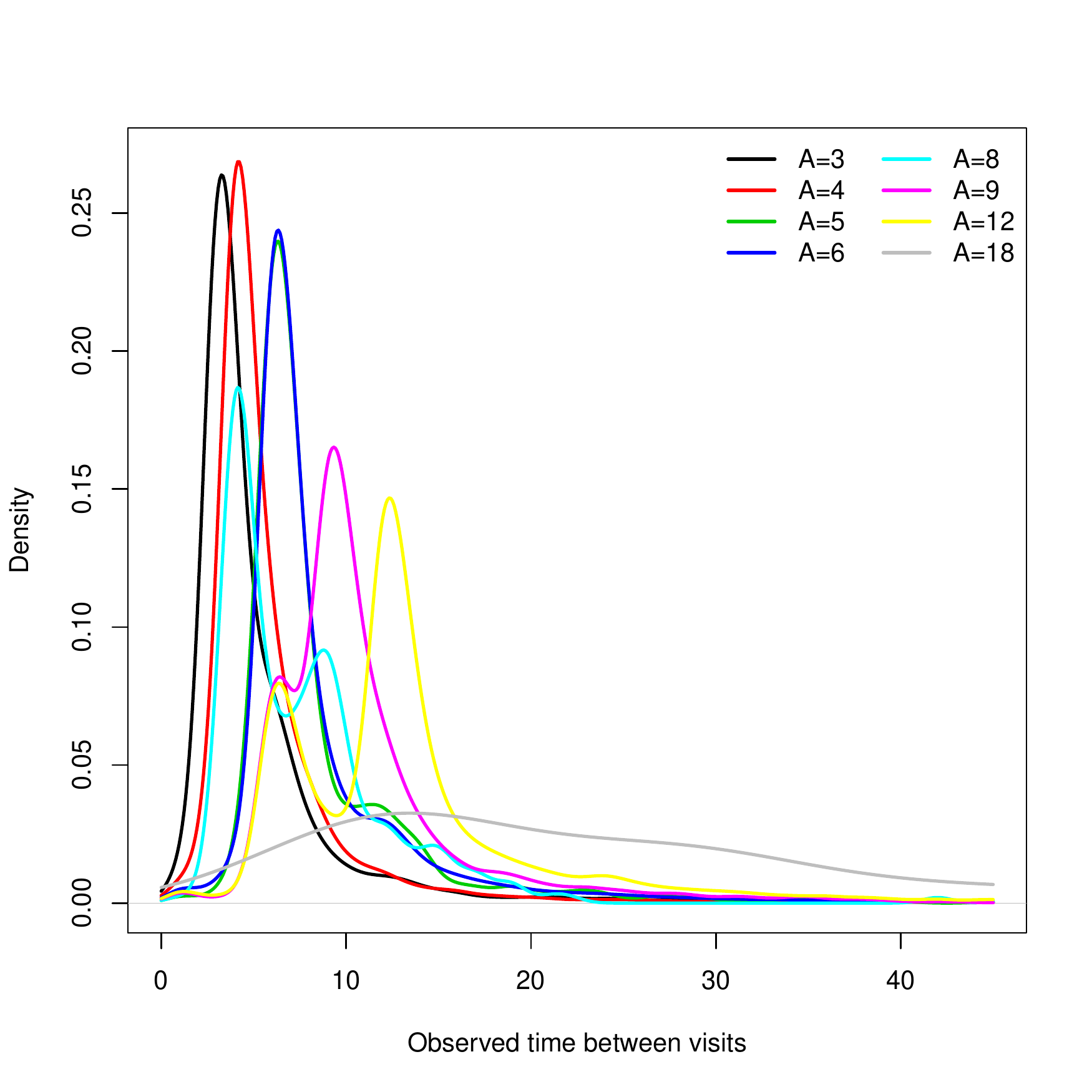}
\end{figure}

Figure \ref{f:subject} plots the longitudinal profiles for 10 subjects. Although there are some short-term decreases, there is a clear population-level increasing trend. A high proportion of the responses are identical to the previous response, reflecting the common practice of carrying the previous values forward in the dental record if there is no apparent change in disease status. During each visit, the recommended time until the next visit ($A$) is also recorded, and the actual time between two visits ($\delta$) is computed.  HP uses an algorithm to classify subjects as low, medium or high risk of PD and caries, and this risk score is taken into consideration when recommending the next visit time. However, this risk score is not optimized for recall recommendations, and dentists are not obliged to use it. The range of $A$, as indicated in Figure \ref{f:density}, varies significantly from 3 months to 18 months. The figure also illustrates the partial controllability, with a strong but imperfect relationship between observed and recommended recall times ($A$) between visits.


\section{Problem statement}\label{s:model}

The data at baseline for subject $i = 1,\ldots, n$ include the $p$-vector of covariates, $\bX_i$ and the scalar baseline response, $Y_{i0}$.
At the baseline visit, a recommended number of months until the first follow-up visit $A_{i1} (> 0)$ is given. The subject returns for the first following visit $\delta_{i1} (> 0)$ months after the baseline visit, and the subject's response, $Y_{i1}$, is recorded. This process is repeated for $N_i$ follow-up visits for the subject $i$. The data available after visit $t$ for subject $i$ are
$\bH_{it} = \{\bX_i,Y_{i0},A_{i1},\delta_{i1},Y_{i1},\ldots,A_{it_i},\delta_{it},Y_{it}\}$,
and $\bH_{i} \equiv \bH_{iN_i}$ is the entire history for subject $i$. The subscript $i$ is suppressed to denote a generic trajectory $\bH_{t} = \{\bX,Y_{0},A_{1},\delta_{1},Y_{1},\ldots,A_{t},\delta_{t},Y_{t}\}$.

Our objective is to use these data to determine a policy for recommending the time between visits.  A policy $\pi$ is a deterministic function that maps the available data to a recommendation, i.e., under $\pi$,
\beq\label{policy}
   A_{t} = \pi(\bH_{t};\balpha).
\eeq
The policy is parameterized in terms of the unknown vector $\balpha = (\alpha_1,\ldots,\alpha_q)^T$. For interpretability, we assume that the policy is a function of a risk score that is a linear combination of features constructed from $\bH_{t}$, $\bff(\bH_{t}) = [f_1(\bH_{t}),\ldots,f_q(\bH_{t})]^T$.  There is great flexibility in constructing the features; they can include the covariates themselves, $f_k(\bH_{t}) = X_{j}$,  or composites such as change from baseline $f_k(\bH_{t}) = Y_{t} - Y_{0}$, or even summaries of the posterior predictive distribution.  The risk score is then $R_t=\bff(\bH_{t})^T\balpha$, and subjects with high risk scores are recommended to have small $A_{t}$, whereas subjects with low risk are recommended to a larger $A_{t}$. While including many features can give a rich class of policies, we consider a small value of $q$ so that the subclass of policies is interpretable, and the optimization problem reduces to estimating the low-dimensional vector $\balpha$. We use $\Pi$ to denote the above pre-specified class of policies that are parameterized by $\balpha$.

We formalize the optimal recall interval recommendation policy within the class $\Pi$ using potential outcomes. Define $Y_{t}^\star(\overline{a}_t)$ and $\delta_{t}^\star(\overline{a}_t)$ to be the potential PMU outcome at visit $t$ and potential time between visit $t$ and visit $t-1$, respectively, if the sequence of recall recommendations $\overline{a}_t$ would be given to a subject since baseline visit, where $\overline{a}_t=\{a_1,\dots,a_t\}$ denotes the history of recall interval recommendations up to visit $t$ . Define $Y_t^\star(\pi)$, $\delta_t^\star(\pi)$ to be the potential outcomes at visit $t$ under recall interval recommendation policy $\pi$. 
The value associated with a policy can be defined as the expectation of a function of potential outcomes under the policy, e.g., $V(\pi)=\mathbb{E}\{[1/J^\star(\pi)]\sum_{t=1}^{J^\star(\pi)}Y_t^\star(\pi)\}$, where $J^\star(\pi)$ is the number of visits within a pre-specified time period under policy $\pi$. The optimal policy within the pre-specified class of policies is $\pi_{opt} = \underset{\pi\in\Pi}{\arg\max}\  V(\pi)$. In order to estimate the optimal recall interval recommendation policy within the pre-specified class using the observed data, we make the following assumptions: (1) no unmeasured confounders (or sequential ignorability) \citep{robins2004optimal}, $\{(Y_{k}^\star(\overline{a}_k),\delta_{k}^\star(\overline{a}_k)):\text{ for all }\overline{a}_k\in\overline{\mathscr{A}_k}\}_{k\geq 1}\perp\!\!\!\perp A_t|\bH_t$ for $t=1,\dots,N$, where $\overline{\mathscr{A}_k}=\mathscr{A}_1\times\dots \times\mathscr{A}_k$ is the set of all possible recall interval recommendations up to visit $k$; (2) consistency, $Y_t=Y^\star(\overline{A}_t)$ and $\delta_t=\delta^\star(\overline{A}_t)$,  where $\overline{A}_t$ is the sequence of observed recommended recall intervals up to visit $t$, i.e., the observed outcomes are the potential outcomes under the actual given recall interval recommendation; (3) positivity, there exists $\epsilon>0$ so that $P(A_t=a_t|\bH_t=h_t)\geq\epsilon$ for all $a_t\in \Psi_t(h_t)$ and for all $h_t$, where $\Psi_t(h_t)$ is the set of possible recall interval recommendations for a subject with realized history information $h_t$, $t=1,\dots,N$. Under those assumptions, our framework of estimating optimal policy is causally interpretable \citep[][]{robins2004optimal,schulte2014q}, and we use the notation of the generic trajectory instead of potential outcomes hereafter.
	

To compare policies, we use a utility function $U(\bH)$,  which is considered the primary outcome to be optimized. Based on the underlying clinical science and logistical constraints, we chose 5-year mean reduction in the proportion of unhealthy sites as one of our primary outcomes; however, the proposed methodology can be extended to other time horizons, or other summaries of a patient’s health trajectory. We desire a policy that applies to all subjects, and we therefore compare the population mean utility, called the value, $V(\balpha) = \mathbb{E}_{\balpha}[U(\bH)]$.
The expectation 
averages over the entire distribution of $\bH$, including the baseline covariates, the visit times as determined by $\delta_{t}$, and the PMU trajectory $Y_0,\ldots,Y_{N}$.   The policy vector $\balpha$ alters the value indirectly via the recommendation times $A_{t}$ which subsequently affect the time courses of $\delta_{t}$ and dental health state $Y_{t}$. Therefore, estimating the value of the policy requires determining compliance relationship (the distribution of $\delta_{t}$ given $A_{t}$), and the effect of recall on PD (the distribution of $Y_{t}$ given $\delta_{t}$). In addition to value, policies must be compared in terms of their cost because it is not feasible to recommend a short time between visits for all subjects.  We control cost by constraining the average recommended recall time to be $T$, $C(\balpha) = \mathbb{E}_{\balpha}(A_{t})=T$.

The objective is to identify an $\balpha$ which maximizes the value $V(\balpha)$ while maintaining cost constraint $C(\balpha)=T$. Rather than attempting to estimate $\balpha$ directly from the data, our approach is to first estimate the distribution of $\bH$ as a function of $\balpha$ using a BNP model (Section \ref{s:BNP}).  Given this model, we can then simulate from the process to obtain Monte Carlo estimates of $V(\balpha)$ and $C(\balpha)$ for any $\balpha$, and use this simulation as a basis for determining the optimal $\balpha$ (Section \ref{s:policy}).

\section{Bayesian model for disease progression}\label{s:BNP}

For our application, we build a Dirichlet Process Mixture (DPM) model that is parsimonious enough to fit large data sets and facilitate the extensive simulation required for policy evaluation, yet flexible enough to capture the complex dynamics of the HP data. Heterogeneity across subjects is captured with subject random effects $\Theta_i = \{\btheta_{i0},\btheta_{i1},\btheta_{i2}\}$ that includes random effects for baseline status ($\btheta_{i0}$), compliance ($\btheta_{i1}$), and disease progression ($\btheta_{i2}$), and is modeled using Bayesian nonparametrics as described below.  Given the random effects, we propose a Markov outcome-dependent follow-up model \citep{ryu2007longitudinal} for $\bH_i$,

\beqn\label{markov}
(\bX_i^T,Y_{i0})^T|\Theta_i &\sim& \mbox{Normal}(\btheta_{i0},\bSigma_0)\\
\log(\delta_{it})|\Theta_i,\bX_i,Y_{it-1},A_{it} &\sim& \mbox{Normal}(\calX_{it}^T\btheta_{i1},\sigma_1^2)\nonumber\\
Y_{it}|\Theta_i,\bX_i,Y_{it-1},\delta_{it} &\sim& \mbox{Normal}(\calZ_{it}^T\btheta_{i2},\sigma_2^2)\nonumber
\eeqn
where $\calX_{it} = [\bX_i^T,Y_{it-1},\log(A_{it}),\bX_i^T\log(A_{it}),Y_{it-1}\log(A_{it})]^T$ and $\calZ_{it} = (\bX_i^T,Y_{it-1},\delta_{it},\bX_i^T\delta_{it},Y_{it-1}\delta_{it})^T$.


Although this first-stage model is relatively simple, the overall model is flexible when integrated over the random effects $\Theta_i$. For example, compliance $\delta_{it}/A_{it}$ depends on both covariates and current disease status, and these relationships are individualized through $\btheta_{i1}$.  Similarly, the individualized treatment effect is controlled by $\btheta_{i2}$ and the induced relationship between $\delta_{it}$, $Y_{it-1}$, and  $Y_{it}$.  Also, prior correlation between $\btheta_{i0}$ and $\btheta_{i2}$ can accommodate effect modification between the baseline covariates and time between visits in the PMU model.  Of course, even more flexible models can be constructed using non-linear terms in $\calX_{it}$ and $\calZ_{it}$ and higher-order lags in the Markov model.

Let $g$ be the random effects density, such that $\Theta_i\iid g(\Theta)$.  Rather than selecting a parametric model for $g$, we treat the density as an unknown quantity to be estimated from the data.  The prior for $g$ is modeled using the Dirichlet process prior \citep{ferguson1973bayesian,sethuraman1994constructive},
which can be written as $g(\Theta) = \sum_{l=1}^L\omega_l\mathbbm{1}_{\Delta_l}(\Theta)$,
where $L=\infty$, the mixture probabilities $\omega_l>0$ satisfy $\sum_{l=1}^\infty\omega_l=1$, 
$\Delta_l=(\btheta_{0l}^{\star T},\btheta^{\star T}_{1l},\btheta_{2l}^{\star T})^T\sim\text{Normal}(\boldsymbol{m}_b,\bSigma_b)$, and $\mathbbm{1}_{\Delta_l}(\cdot)$ is the indicator function with a  point mass at $\Delta_l$. The mixture probabilities can be generated from the stick-breaking process: $\omega_l=V_l\prod_{h<l}^{}(1-V_h)$, $V_l\sim \text{Beta}(1,\alpha_0)$.
The covariance matrix $\bSigma_b$ is taken to be block diagonal with $\mbox{Cov}(\btheta_{jl}^\star) = \bSigma_{bj}$ and $\mbox{Cov}(\btheta_{jl}^\star,\btheta_{kl}^\star) = 0$.  For priors, we select  $\boldsymbol{m}_b\sim\mbox{Normal}(0,I)$ and $\bSigma_{bj}\sim\mbox{InvWishart}(p_j+1,(p_j+1)I_{p_j})$, where $p_j$ is the dimension of $\btheta^\star_{jl}$. With these priors and truncation at a finite $L$, all full conditional distributions are conjugate and so we use Gibbs sampling to obtain posterior samples as described in the supplemental material.


\section{Policy search}\label{s:policy}

\begin{figure}
	\caption{\small{Illustration of how to optimize a sequence of recall intervals under a given policy using the proposed method. In this hypothetical example, the risk score $R_t$ is a linear combination of the current disease statue ($Y_t$) and a single covariate (X), and the two actions are to return in 3 or 9 months.}}\label{f:illus}
	\centering
	\includegraphics[height=0.5\textwidth]{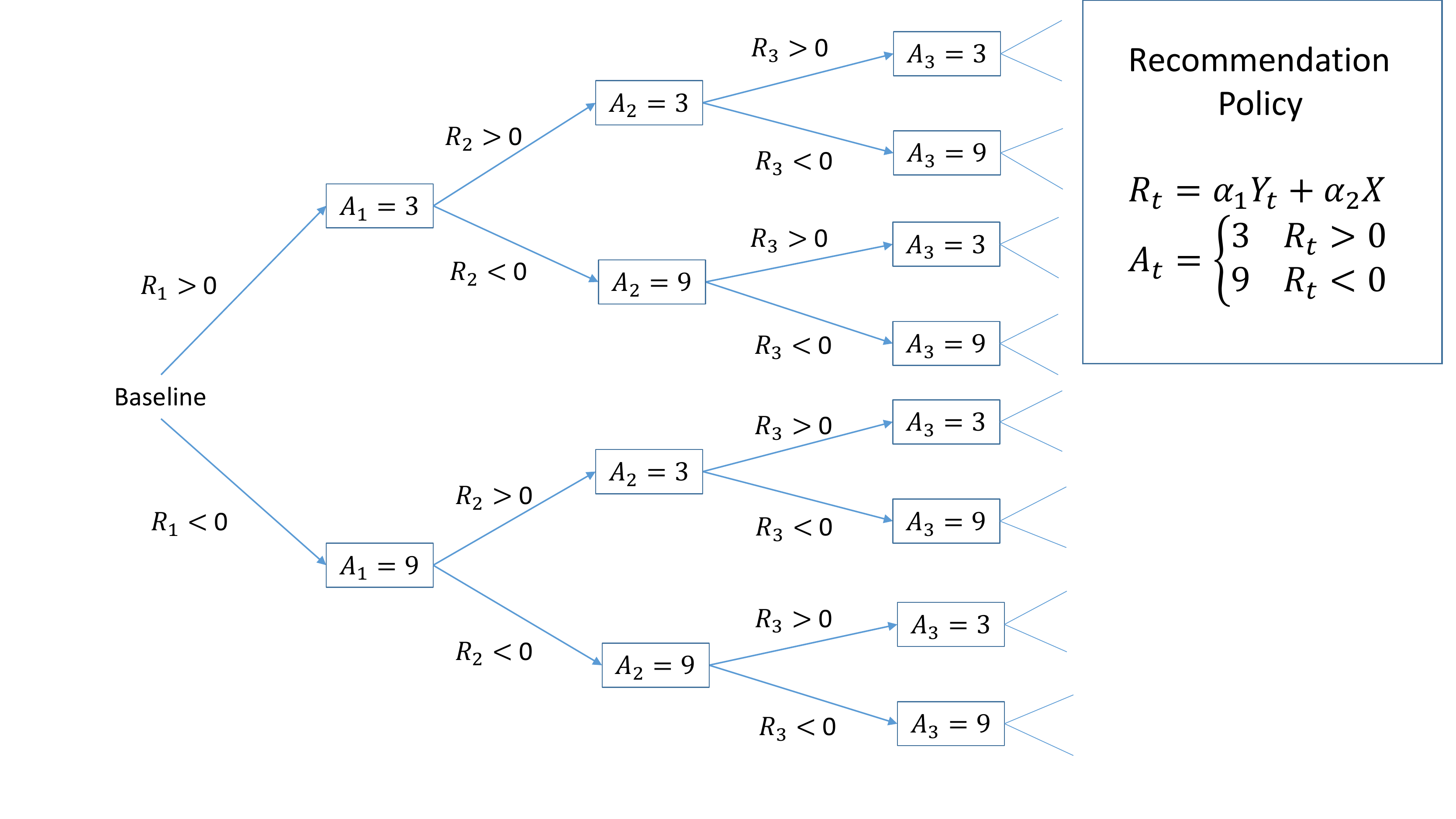}
\end{figure}

Although other classes of policies are possible, such as trees \citep{laber2015tree} and lists \citep{zhang2015using}, we consider policies defined by a linear risk score. Let the risk score be $R_{t} = \bff(\bH_{t})^T\balpha$, where $\bff(\bH_{t})$ is a $q$-vector of features and $\balpha=(\alpha_1,\ldots,\alpha_q)^T$ are their weights. Our general framework can easily accommodate non-linear relationships between patient characteristics and the risk score by including non-linear summaries of the characteristics as features. As an extreme example, we could include B-spline basis or tree expansion of a variable as features to give an arbitrarily flexible risk score. We could also include an interaction between a characteristic and disease status to account for different importance of the characteristic as the disease progresses. However, our goal is to develop a policy that is interpretable to domain experts so that it can be integrated to the clinical practice. Hence, we decided to keep the risk score simple. We describe the method assuming two possible recommendations, $a_1$ and $a_2$. The policy takes the form
\beq\label{linearpolicy}
A_{t} = \pi(\bH_{t};\balpha) =
  \begin{cases} a_1 &  R_{t}>\kappa(\balpha) \\
	a_2 & R_{t} \le \kappa(\balpha) \end{cases}
\eeq
where $\kappa(\balpha)$ is the risk threshold that depends on $\balpha$; more than two treatments could be accommodated using multiple thresholds.  With only a single threshold, the scale of $\balpha$ is irrelevant, and so we impose the restriction $||\balpha|| = (\sum_{j=1}^q\alpha_j^2)^{1/2}=1$. Figure \ref{f:illus} illustrates how the proposed method is used when a sequence of recall intervals needs to be optimized under a given policy in terms of $\balpha$.

We select the policy parameters $\balpha$ so that when the population of patients follows this rule over time, the long-term average reward is high. Given the posterior of the random effects distribution $g$ and covariance parameters $\calS = \{\Sigma,\sigma_1, \sigma_2\}$, the optimal feature weight $\balpha$ is given by
\beq\label{optimal}
  \balpha_{opt} = \underset{\balpha}{\arg\max}\  V(\balpha) \mbox{\ \ such that \ \ } C(\balpha)=T.
\eeq
However, this is a challenging optimization problem, because both $V(\balpha)$ and $C(\balpha)$ are expectations with respect to the predictive distribution of $\bH$ given $g$ and $\calS$. Both $V(\balpha)$ and $C(\balpha)$ are approximated using Monte Carlo sampling.  The MCMC algorithm described in Section \ref{s:BNP} produces $J$ posterior draws $\{g^{(j)}, \calS^{(j)}; j=1,\ldots,J\}$. For each candidate $\balpha$ and $\kappa(\balpha)$, we simulate subject $i=1,\ldots,n_0$ by first sampling $j$ randomly from $\{1,\ldots,J\}$, then $\Theta_i\sim g^{(j)}$, and finally $\bH_i$ from (\ref{markov}) given $\Theta_i$ and $\calS^{(j)}$, with recommendations given by $\pi(\bH_{it};\balpha)$. Note that in these simulations, the actions taken affect the future outcomes \& thus future recommendtaions, and in this way the method can allow for delayed effects.

For each candidate $\balpha$, we first identify the threshold $\kappa(\balpha)$ that satisfies the cost constraint $C(\balpha)\approx T$, and then estimate the value $V(\balpha)$ given this threshold.  We estimate $C(\balpha)$ separately for a grid of 10 thresholds spanning the range of $R_{t} $ in the training data, smooth the estimates using LOESS \citep{cleveland1988locally}, and then compute the $\kappa$ that gives $C(\balpha)\approx T$.  For each of the 10 candidate thresholds, we draw $n_1=2,000$ independent subjects' $\bH_i$ and use the sample mean of the $n_1$ averaged recommended recall time as the estimate of $C(\balpha$).  Given the threshold $\kappa(\balpha)$, the value $V(\balpha)$ is approximated by another round of Monte Carlo simulation of $n_2=20,000$ subjects.

We utilize response-surface sequential optimization methodology \citep{mason2003statistical} to identify the value of $\balpha$ that maximizes $V(\balpha)$.  The objective function of this optimization is noisy because only a Monte Carlo estimate of the value is available. In the first stage, we evaluate the value using a central composite design for $\balpha$ \citep{montgomery2008design}, scaled to $||\balpha||=1$.  That is, we consider the $M_1=5^q-1$ unconstrained values ${\tilde \balpha}\in \{-2,-1,0,1,2\}^q$ (excluding the zero vector), and then the corresponding constrained  vectors $\balpha_l = (\alpha_{l1},\ldots,\alpha_{lq})^T$ formed by setting $\alpha_{lj} = \tilde{\alpha}_{lj}/\sqrt{\sum_{h=1}^q{\tilde \alpha}_{lh}^2}$, such that $||\balpha_l||=1$.  A Monte Carlo estimate of the value ${\hat V}_l$ is computed for each $\balpha_l$, giving $M_1$ pairs $\{\balpha_l,{\hat V}_l\}$.

For a given $\balpha$, the value can be estimated using extensive simulation from the DPM model. However, when searching for the next $\balpha$ to consider, we need a quick approximation to the value to avoid spending too much time simulating the value for poor policies. Using these training data, we fit a Gaussian process regression model to quickly predict the value of a new policy (via $\balpha$), and guide the remaining optimization steps. The value is modeled as a Gaussian process with $\mbox{E}({\hat V}_l) = \mu_V$, $\mbox{Var}({\hat V}_l) = \sigma_V^2$, and correlation $\mbox{Cor}({\hat V}_l,{\hat V}_k) = (1-r)\mbox{I}(l=k) + r\exp[-\sum_{j=1}^q\phi_j(\alpha_{lj}-\alpha_{kj})^2]$.  We set $\mu_V$ and $\sigma_V^2$ to the sample mean and variance of ${\hat V}_l$, respectively.  The correlation parameter $r$ is set to 0.99; the maximum likelihood estimates of $r$ were near one which led to computational issues with singular covariance matrices, that where alleviated by setting $r=0.99$ \citep{gramacy2012cases}. With these parameters fixed, we compute maximum likelihood estimates $\phi_1,\ldots,\phi_q$.

We simulate the values ${\hat V}_{M_1+1}$,\ldots,${\hat V}_{M_1+M_2}$ corresponding to an additional $M_2=200$ feature weights $\balpha_{M_1+1}$,\ldots,$\balpha_{M_1+M_2}$ using the sequential optimization criteria of \cite{jones1998efficient}. The policy weights at step $l>M_1$ are selected to optimize the expected gain in the optimal value.  Let ${\tilde V}_l = \max\{{\hat V}_1,\ldots,{\tilde V}_{l-1}\}$ be the maximum value observed prior to step $l$, and define the expected increase in the maximum value if we take an additional sample at $\balpha$ as $$G(\balpha) = \Phi\left[\frac{m(\balpha)-{\tilde V}_l}{s(\balpha)}\right]\left[m(\balpha)-{\tilde V}_l\right] + s(\balpha)\phi\left[\frac{m(\balpha)-{\tilde V}_l}{s(\balpha)}\right],$$ where $m(\balpha)$ and $s(\balpha)$ are the predictive mean and standard deviation of ${\hat V}$ at $\balpha$ from the Gaussian process regression model using the first $l-1$ observations, and $\Phi$ and $\phi$ are the standard normal distribution and density functions. To approximate the maximizer of $G(\balpha)$, we randomly generate 1,000 $\balpha$, scale them so $||\balpha||=1$, compute $G(\balpha)$ for each $\balpha$, and take $\balpha_l$ to be the $\balpha$ with the largest $G(\balpha)$. The final estimate is the $\balpha$ that maximizes $m(\balpha)$, the predictive mean from Gaussian process regression given the $M_1+M_2$ training points. This optimization is approximated by sampling $M_3=20,000$ weights $\balpha_{1},\ldots,\balpha_{M_3}$, scaling them so that $||\balpha||=1$, and computing
\beq\label{ahat}
\balpha_{opt} \approx \underset{\balpha\in\{\balpha'_1,\ldots,\balpha'_{M_3}\}}{\operatorname{\argmax}}m(\balpha).
\eeq

With these specifications, the optimization requires approximately 75 minutes on a standard desktop computer for the simulated cases described in Section \ref{s:sim}.  However, this rudimentary {\it R} code does not exploit the obvious opportunities to parallelize over the subjects within the Monte Carlo simulations for a given $\balpha$ or across simulations for different $\balpha$.  Therefore, it should be possible to scale this approach up to larger problems than those considered here. The \texttt{R} package \texttt{DiceOptim} \citep{roustant2012dicekriging} performs stochastic optimization using similar steps as our algorithm, so users may be able to use this package to avoid extensive coding for some of the proposed optimization steps.

MCMC produces posterior draws of the random effects distribution and covariance parameters, $f^{(s)}$ and $\calS^{(s)}$, for $s=1,\ldots,S$ MCMC samples.  Each posterior sample corresponds to a different ${\balpha}_{opt}$.  Applying this optimization for each posterior draws produces a posterior distribution for ${\balpha}_{opt}$, which is used for uncertainty quantification.

\section{Simulation study}\label{s:sim}


Each dataset consists of $n$ subjects generated independently from (\ref{markov}).  There are $p=2$ baseline covariates, the variance parameters are  $\sigma_1=0.1$ and $\sigma_2=0.5$, and $\bSigma_0$ is the correlation matrix with 0.5 for all off-diagonal elements. Subjects are generated from two groups, with the group identifier for subject $i$ denoted $G_i\in\{1,2\}$. Subjects from the first group are generated as

\beqn\label{markov1}
(\bX_i^T,Y_{i0})^T|G_i=1 &\sim& \mbox{Normal}(0,\bSigma_0)\\
\log(\delta_{it})|Y_{it-1},A_{it},G_i=1 &\sim& \mbox{Normal}\left[\log(A_{it})(0.9+0.1X_{i1}),\sigma_1^2\right]\nonumber\\
Y_{it}|Y_{it-1},\delta_{it},G_i=1 &\sim& \mbox{Normal}\left[0.1+0.2X_{i2}+0.2(\delta_{it}-6) + 0.9Y_{it-1}+0.02(\delta_{it}-6) Y_{it-1},\sigma_2^2\right].\nonumber
\eeqn
Subjects from the second group are generated as
\beqn\label{markov2}
(\bX_i^T,Y_{i0})^T|G_i=2 &\sim& \mbox{Normal}\left[(1,0,0)^T,\bSigma_0\right]\\
\log(\delta_{it})|Y_{it-1},A_{it},G_i=2 &\sim& \mbox{Normal}\left[\log(5.3),\sigma_1^2\right]\nonumber\\
Y_{it}|Y_{it-1},(\delta_{it}-6),G_i=2 &\sim& \mbox{Normal}\left[0.1+0.3X_{i1}-0.2(\delta_{it}-6) + 0.9Y_{it-1},\sigma_2^2\right].\nonumber
\eeqn
Unlike the first group, the second group of subjects are non-compliers in that the recommendation $A_{it}$ does not affect the distribution of the time until next visit.  For simulated training data, recommendations are either $A_{it}\in\{3,9\}$ with $\mbox{logit}[\mbox{Prob}(A_{it}=3)]=Y_{it-1}$.  For each subject, we simulate observations until the subject has been in the study for five years. We consider two scenarios by varying the cluster assignment probability. The cluster assignment is either ``Single group'' with $\mbox{Prob}(G_i=1) = 1$ or ``Mixture model'' with $\mbox{Prob}(G_i=1) = 0.8$. The sample size is $n=1,000$. We simulate 100 datasets from each scenario. The supplemental materials include additional simulations with binary covariates and misspecified models.


We consider two different utility functions: $U(\bH) = -1/N\sum_{t=1}^{N}Y_{t}$ (``average''), which focuses the policy to reduce large values of $Y_{t}$; $U(\bH) = Y_{0}-Y_{T_{60}}$ (``reduction''), which aims to maximize the reduction of PMU in 5 years from baseline, where $Y_{T_{60}}$ is the response for a subject in 5 years (60 months) since baseline visit (if no visit occurs at the exact time point, interpolation is used to estimate the response value). We compare four methods. The ``baseline'' policy recommends $A_{t}=6$ months between visits for all subjects and $t$.  The remaining three methods use the policy in (\ref{linearpolicy}) with $a_1=3$ and $a_2=9$. The risk score is a linear combination of $q=4$ features representing the two baseline covariates ($X_{1}$ and $X_{2}$), non-compliance (via $\log(|\delta_{t-1}-A_{t-1}|+1)$), and disease status (via $Y_{t-1}$): $$R_{t}=X_{1}\alpha_1+X_{2}\alpha_2+\log(|\delta_{t-1}-A_{t-1}|+1)\alpha_3 + Y_{t-1}\alpha_4$$ with $A_{0}=\delta_{0}=6$. We compare two methods that estimate $\balpha$ and $\kappa(\balpha)$ by fitting the $n$ training observations with either a ``Gaussian'' model or ``DPM'' model, and then approximating the value using the posteriors of $g$ and $\calS$ as described in Section \ref{s:policy}.  For the DPM model, we use $L=5$ mixture components, and for the Gaussian model, we use the DPM model with $L=1$.  We also compare the ``oracle'' policy which computes $\balpha$ and $\kappa(\balpha)$ by simulation assuming the true values of the model parameters in (\ref{markov1}) and (\ref{markov2}).  Of course, in a real data analysis, this would be impossible, but we include this in the simulation study as a reference. The hyperparameter values and MCMC details are described in the supplemental material.

The Gaussian and DPM models are fitted to the data using MCMC sampling with $J=5000$ iterations. For these methods, the policy via $\balpha$ and $\kappa(\balpha)$ is computed using Monte Carlo simulation given posterior samples, using the fit to the training data.  After estimating the policy, the averaged recommended recall time and value of these methods are approximated using sample means over 1,000,000 Monte Carlo draws, assuming the true parameter values in (\ref{markov1}) and (\ref{markov2}). For each of the 100 simulated datasets, we estimate one optimal policy and the value corresponding to the estimated optimal policy for each utility function. Table \ref{t:sim} reports the mean of the 100 values and average recommended recall time over the 100 simulated datasets for each scenario and each utility function. Since the standard error is bounded by 0.01, we present the sample means by rounding them to two decimal places. For the baseline policy, there are no policy parameters to be estimated, hence the value is simply approximated using sample means over 1,000,000 Monte Carlo draws given the true parameter values.  The oracle model requires estimating $\balpha$ and $\kappa(\balpha)$, but the estimates $\balpha$ and $\kappa(\balpha)$ do not depend on the training data.


All three adaptive policies have larger (better) value than the static baseline policy in all cases.  For data generated from a single group, the Gaussian model is correct and produces value nearly identical to the oracle policy.  The DPM approach is also nearly identical to the oracle model in this case, showing that little is lost in fitting a complex model in this simple case.  When data are generated from the two-component mixture model that includes non-compliers, the misspecified Gaussian model gives a policy with suboptimal value and averaged recommended recall time that exceeds the six-month threshold. For the mixture model, the DPM approach provides a substantial improvement over the Gaussian model.

\begin{table}\caption{\small{Simulation study results comparing the baseline model with 6-month recommendation for all subjects, policy search methods based on Gaussian and Dirichlet process mixture (DPM) fits, and the oracle model which uses the true data-generating model to estimate the policy. The standard errors of the sample means are all less than 0.01.
	}}
	\label{t:sim}
	\begin{center}\begin{tabular}{llcccc}
			Cluster & Utility &\multicolumn{4}{c}{{\bf Value} (larger is preferred)}\\
			Allocation & Function & Baseline & Gaussian & DPM & Oracle \\ \hline
			Single & Average &  -0.67 & -0.09  & -0.09  & -0.09 \\
			Single & Reduction &  -0.87 & -0.05 & -0.05 & -0.05 \\
			Mixture & Average &  -1.01 & -0.63 & -0.58 & -0.58 \\
			Mixture & Reduction &  -1.42 & -0.86  & -0.83  & -0.83 \\
			& & & & \\
			Cluster & Utility&\multicolumn{4}{c}{{\bf Average recommended recall time}}\\
			Allocation & Function& Baseline & Gaussian & DPM & Oracle \\ \hline
			Single & Average &  6.00 & 5.99  & 5.99  & 6.00 \\
			Single & Reduction &  6.00 & 5.99  & 5.99 & 6.00 \\
			Mixture & Average &  6.00 & 6.07 & 5.99 & 6.00 \\
			Mixture & Reduction &  6.00 & 6.07 & 5.99 & 6.00 \\
		\end{tabular} \end{center}\end{table}

\begin{figure}
	\caption{\small{Estimated optimal feature weights $\balpha_{opt}$ for the simulation study.  The boxplots for the Gaussian and DPM methods show the estimated $\balpha_{opt}$ over the 100 simulated datasets; the solid points represent $\balpha_{opt}$ for the Oracle policy.  The risk score is a linear combination the two baseline covariates ($X_{1}$ and $X_{2}$), non-compliance (``Non-comp''; $\log(|\delta_{t-1}-A_{t-1}|+1)$), and disease status (``Cur Y'';  $Y_{t-1}$):}}\label{f:alpha_sim}
	\begin{center}\begin{picture}(420,220)
		\includegraphics[height=0.48\textwidth]{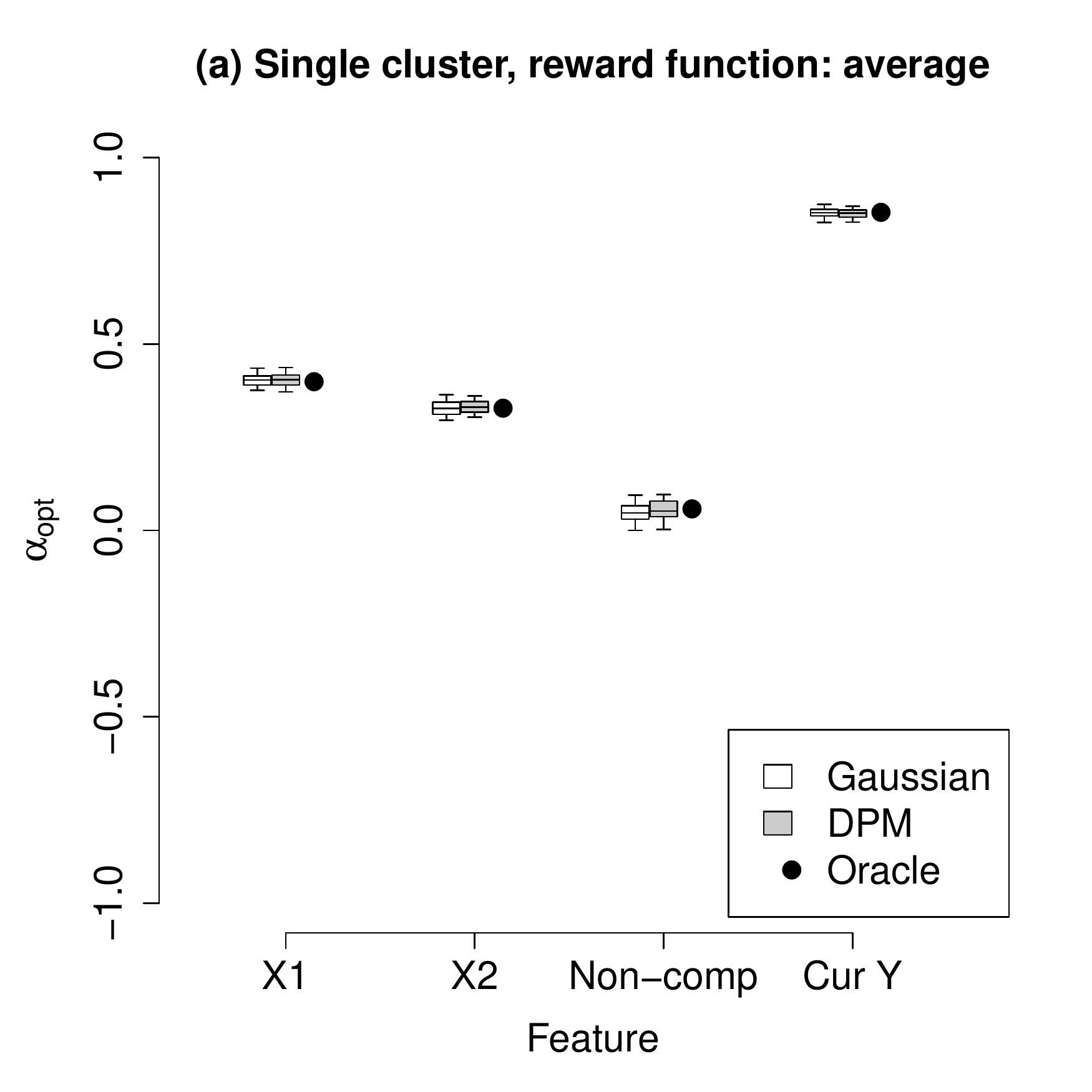}
		\includegraphics[height=0.48\textwidth]{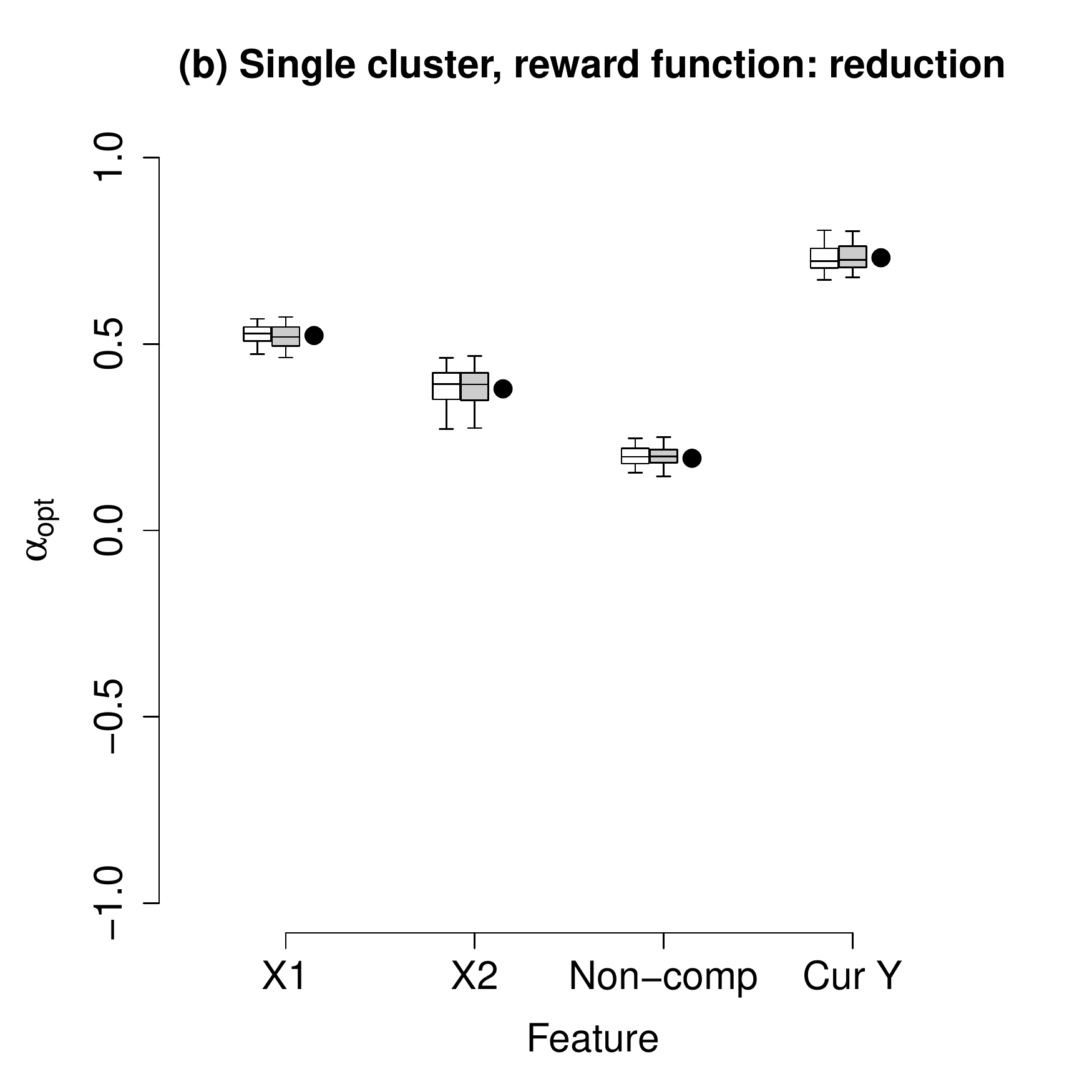}
		\end{picture}
	    \begin{picture}(420,220)
		\includegraphics[height=0.48\textwidth]{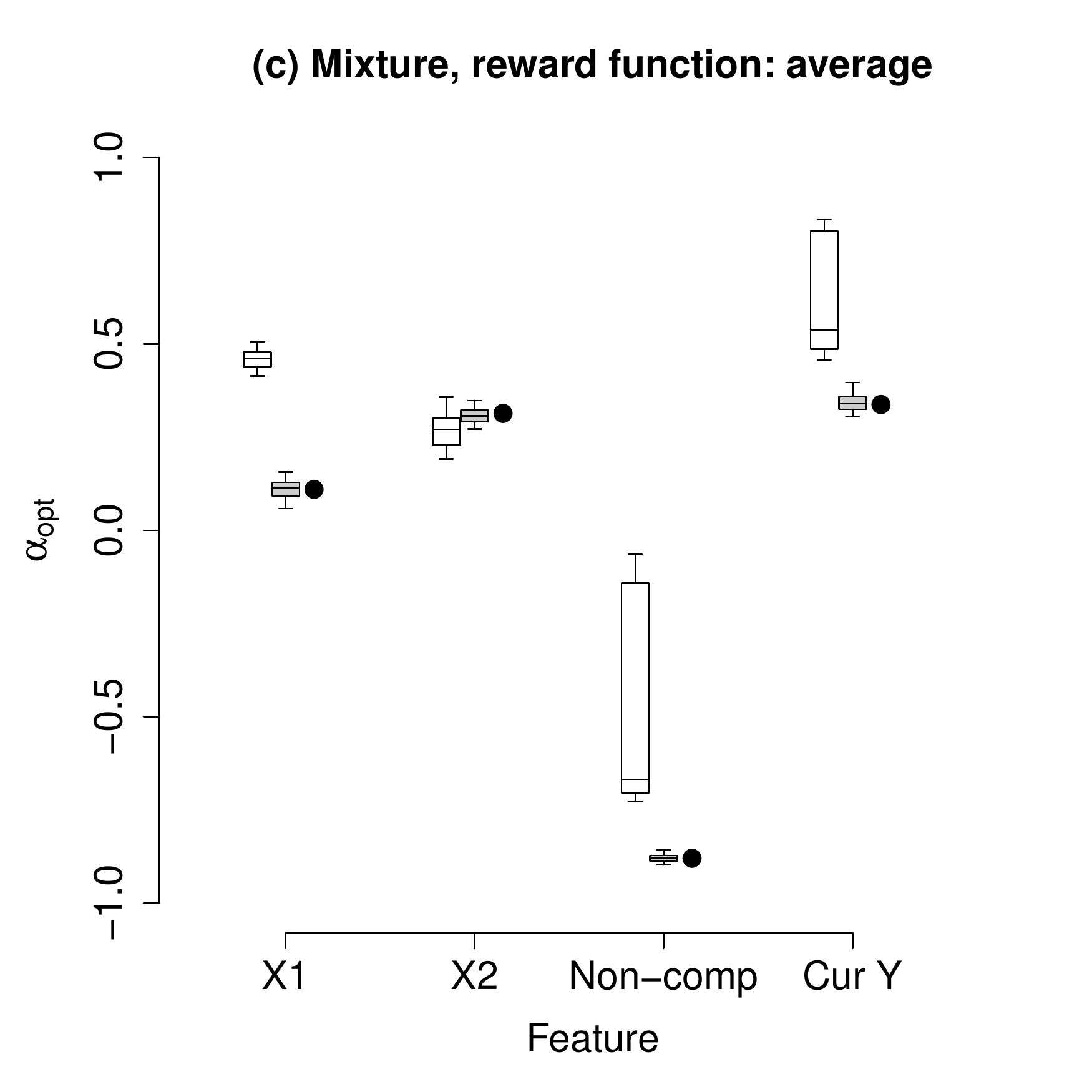}
		\includegraphics[height=0.48\textwidth]{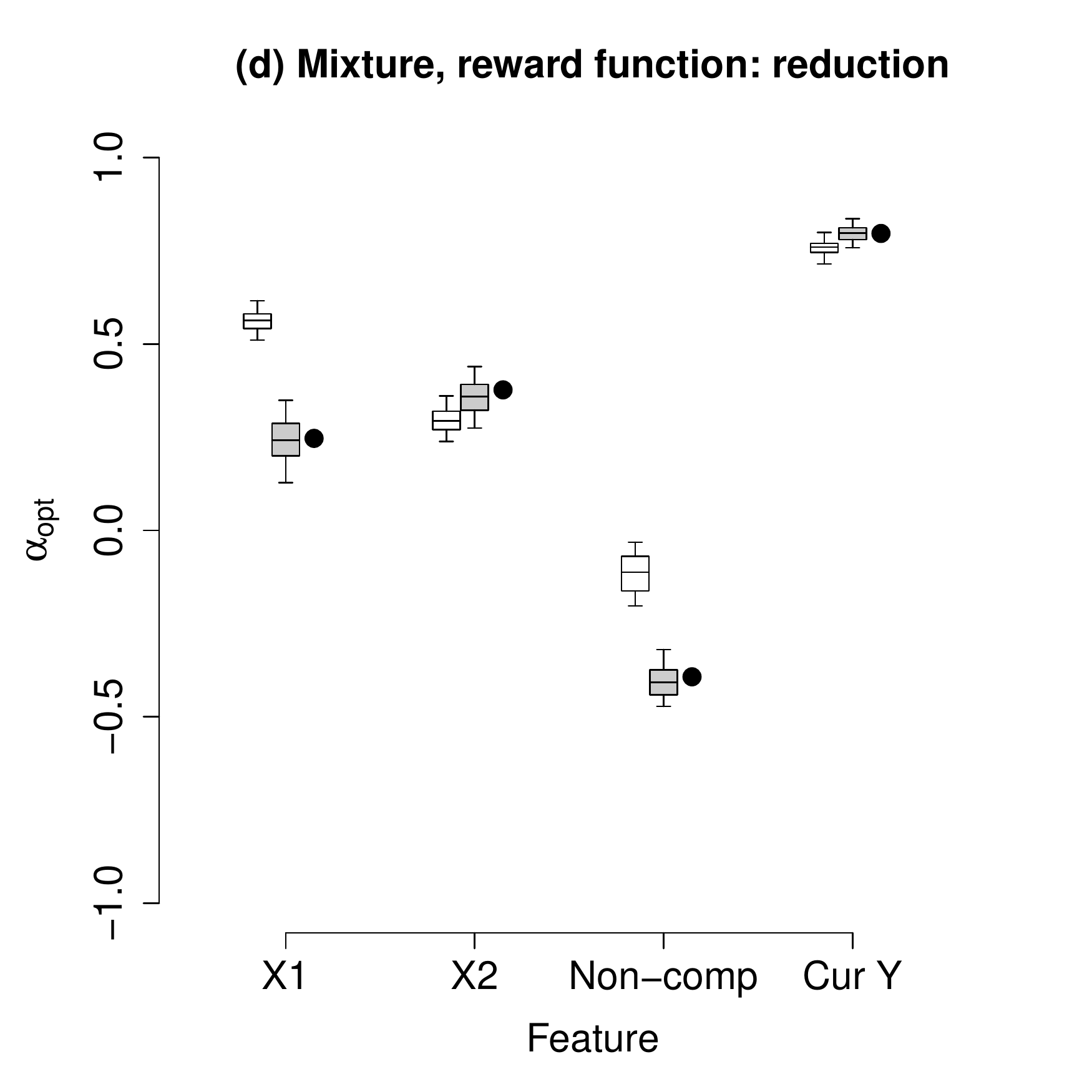}
		\end{picture}\end{center}
\end{figure}

To gain further insight about the effects of model misspecification, Figure \ref{f:alpha_sim} plots the sampling distribution of the estimated policy weights $\balpha_{opt}$ for each method, scenario and utility function. Both the Gaussian and DPM methods give $\balpha_{opt}$ near the oracle model for the single-group scenario.  In this case, the previous value of $Y$ is the most important feature and thus the policy is to recommend shorter recall times for unhealthy subjects. The estimated $\balpha_{opt}$ under the Gaussian model disagree with the oracle policy for data generated under the mixture model.  For example, the importance of non-compliance is underestimated.  In contrast, the oracle model in Figure \ref{f:alpha_sim} (c) gives considerable weight to non-compliance to account for non-compliers.

Model misspecification for the Gaussian case also affects the estimated value and average recommended recall time of the policy. Figure \ref{f:value_sim} shows that the value is generally larger, thus overly optimistic when evaluated using Monte Carlo simulations under the incorrectly fitted model than the true mixture model. In practice, value must be estimated under the fitted model, which can be misleading if the model is incorrect.


\begin{figure}
	\caption{\small{Value and average recommended recall time for the 100 datasets for the policy (via $\balpha_{opt}$) estimated using the Gaussian model or DPM model for data generated from a two-component mixture model with $n=1,000$ subjects using reduction utility function. The value and average recommended recall time of the policy are evaluated using Monte Carlo samples under the true model used to generate the data and the estimated Gaussian model or DPM model.}}\label{f:value_sim}
	\subcaption{Gaussian model}
	\begin{center}\begin{picture}(420,220)
		\includegraphics[height=0.48\textwidth]{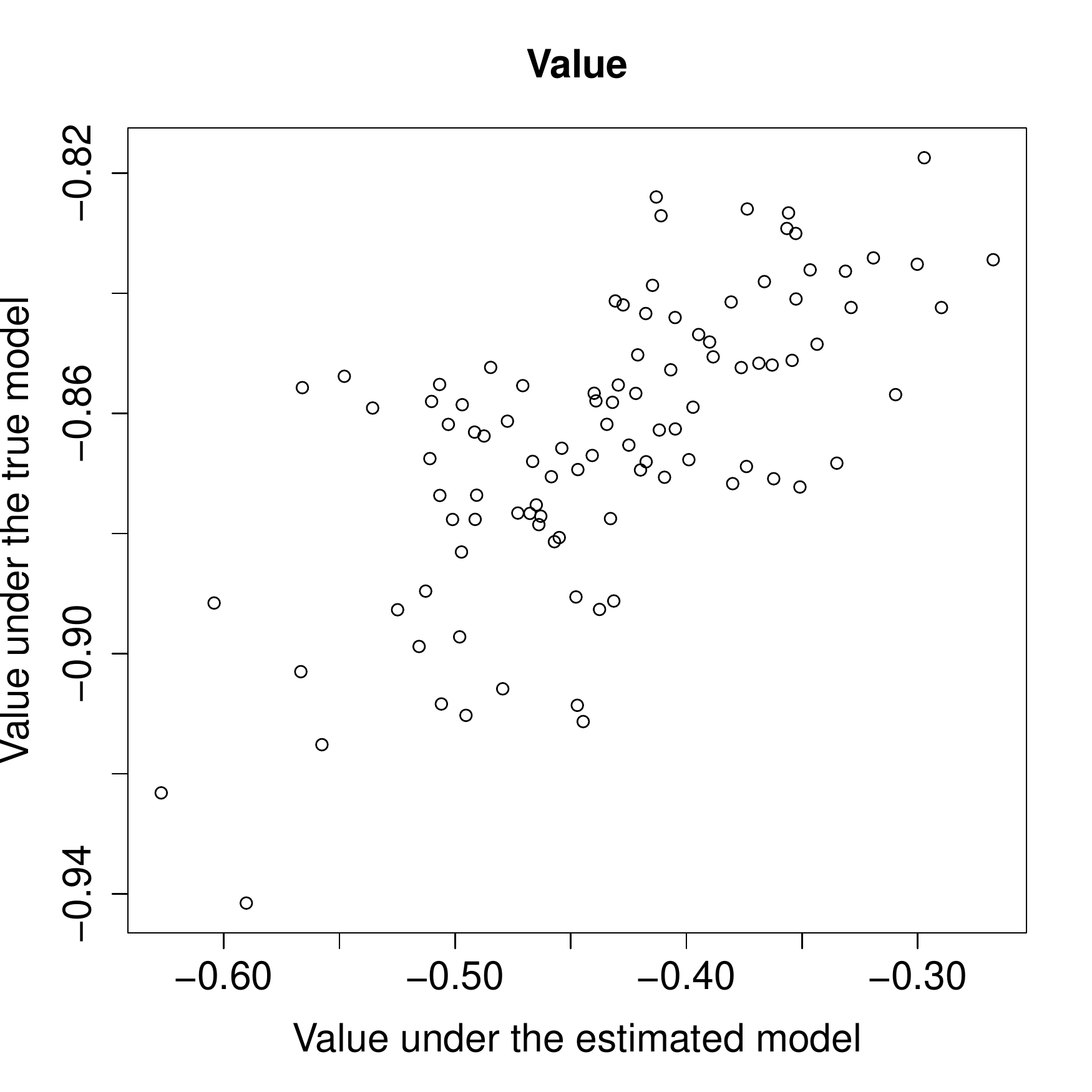}
		\includegraphics[height=0.48\textwidth]{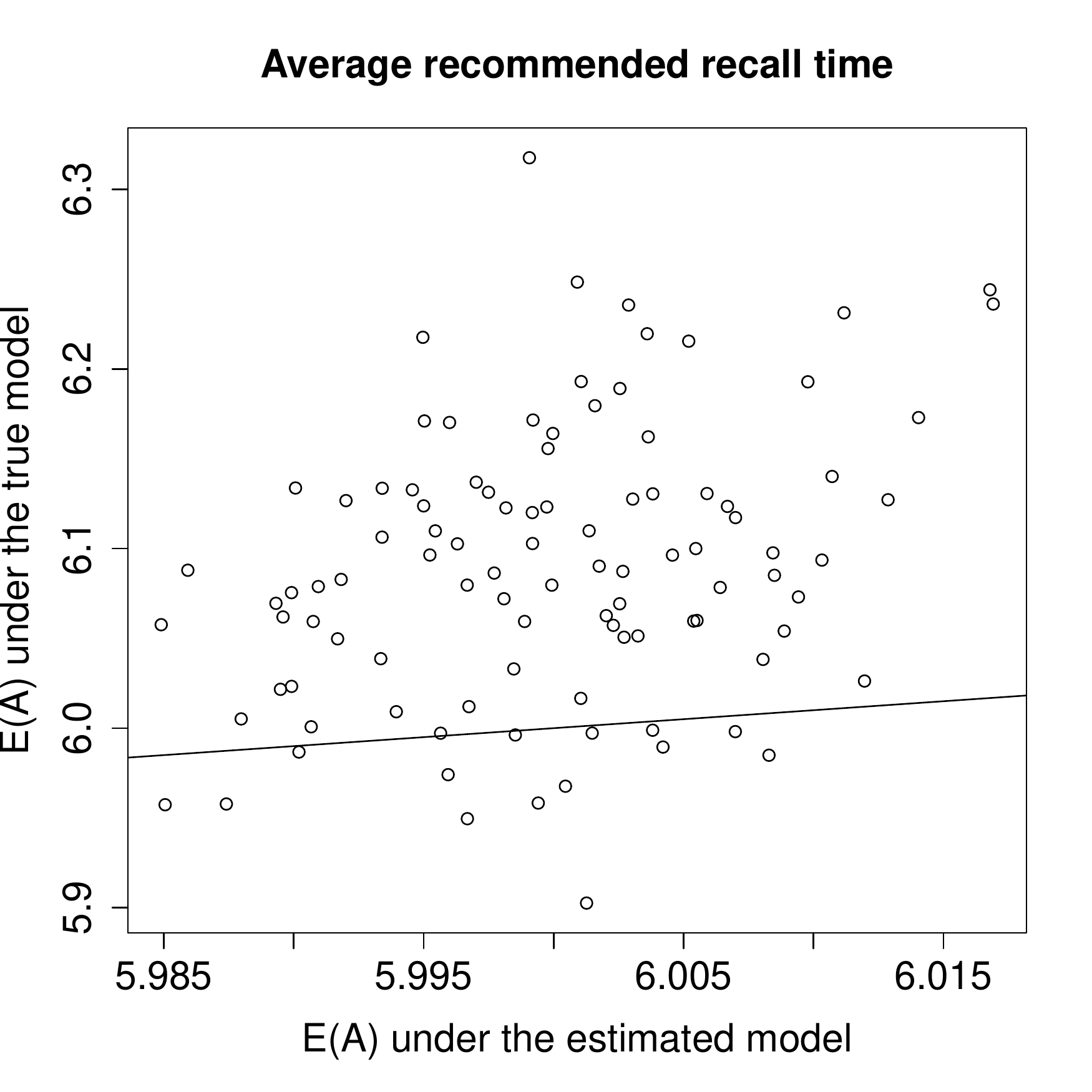}
	\end{picture}\end{center}\end{figure}

\begin{figure}
	\ContinuedFloat
	\subcaption{\small{DPM model}}
	\begin{center}\begin{picture}(420,220)
		\includegraphics[height=0.48\textwidth]{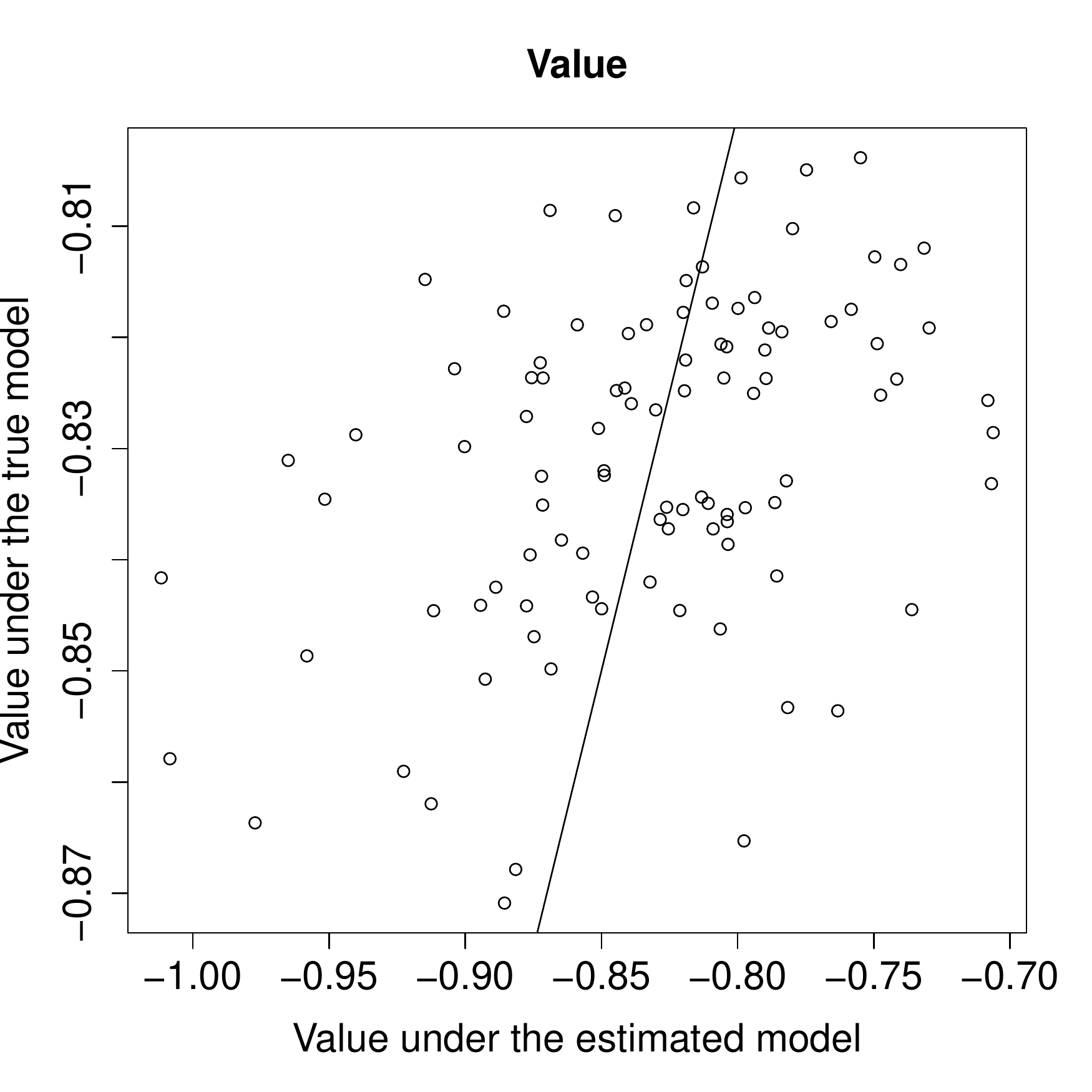}
		\includegraphics[height=0.48\textwidth]{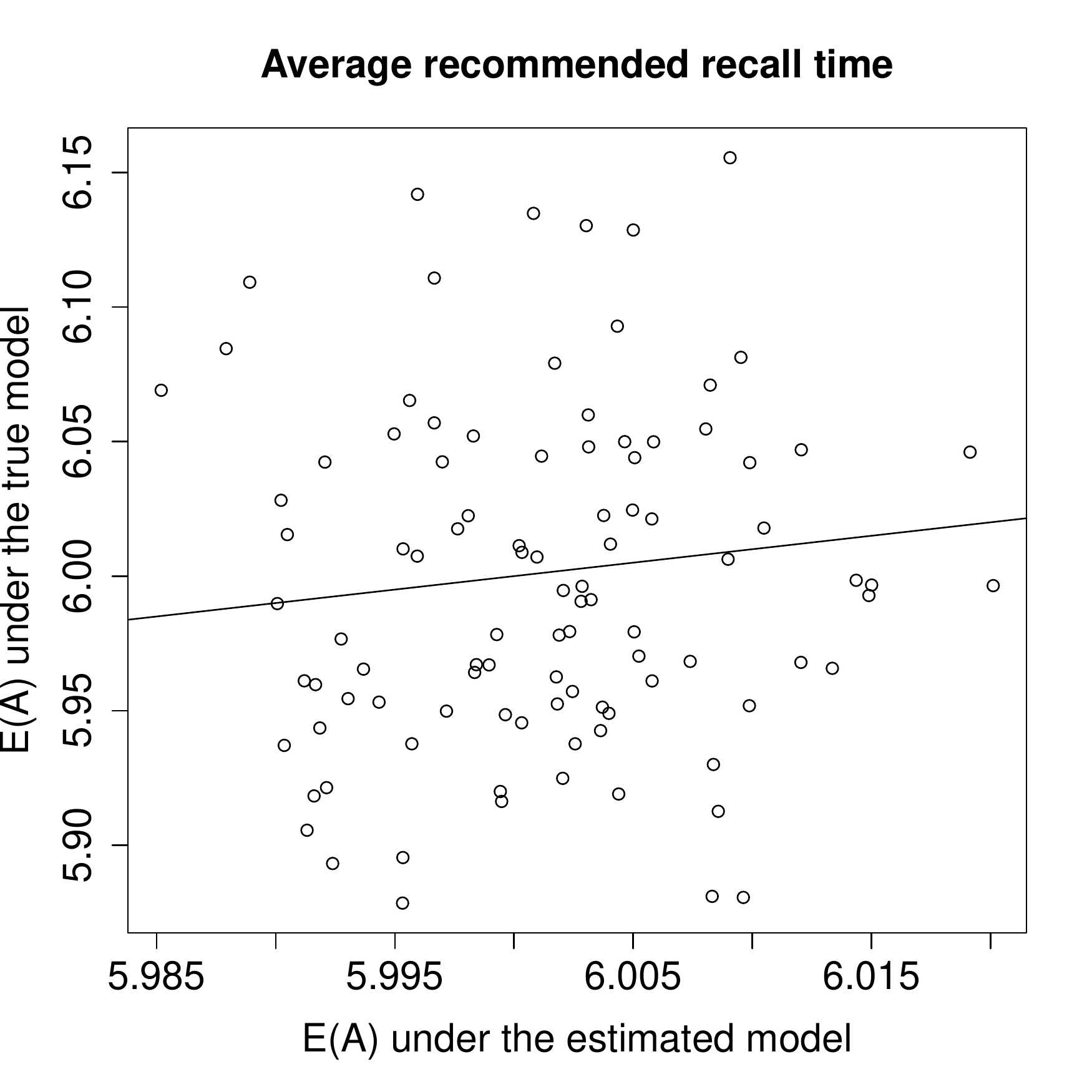}
		\end{picture}\end{center}\end{figure}

The optimal $\balpha$ in (\ref{optimal}) is a deterministic function of the model parameters $g$ and $\calS$. Thus far, we have been averaging over uncertainty in $g$ and $\calS$ to obtain an optimal policy. However, to quantify uncertainty in the policy, we can inspect the posterior distribution of $\balpha_{opt}$ induced by the posterior distribution of $g$ and $\calS$. To illustrate, we simulate 100 datasets generated from the two-component mixture model. We randomly select posterior samples $\{g^{(j)}, \calS^{(j)}; j=1,\ldots,20\}$ from all $J$ posterior draws produced by fitting the DPM model to the data. Given each selected posterior sample of model parameters, a posterior $\balpha_{opt}$ is estimated using reduction utility function. We estimate 90\% credible intervals of $\balpha_{opt}$ for each of the 100 simulated datasets. More specifically, for each simulated dataset, we estimate the optimal policy corresponding to each of the posterior draw of $g$ and $\calS$ to obtain the posterior distribution of the optimal weight $\balpha_{opt}$. The coverage rates of credible intervals for optimal $\alpha_1$, $\alpha_2$, $\alpha_3$, $\alpha_4$ are 96\%, 95\%, 98\%, 100\% respectively, which indicates that our estimated credible intervals are conservative and reliable.

\section{Analysis of the HealthPartners data}\label{s:HP}

\subsection{Tailoring the BNP model to the HP data}\label{s:adjustment}
The DPM model in (\ref{markov}) must be generalized to incorporate the complexities of the HP data. In the HP data, the baseline covariate vector $\bX_i$ includes gender $(X_{i1})$, race $(X_{i2})$, standardized age $(X_{i3})$, diabetes status $(X_{i4})$, smoking status $(X_{i5})$ and commercial insurance indicator $(X_{i6})$. In the final model, we include covariates, such that $\calX_{it} = [\bX_i^T,Y_{it-1},\log(A_{it}),\bX_i^T\log(A_{it}),Y_{it-1}\log(A_{it})]^T$ and $\calZ_{it} = [\bX_i^T,Y_{it-1},\log(\delta_{it}),\bX_i^T\log(\delta_{it}),Y_{it-1}\log(\delta_{it})]^T$. Because $X_{i1}$, $X_{i2}$, $X_{i4}$, $X_{i5}$ and $X_{i6}$ are binary, we introduce latent continuous variables $\bX_i^\star$ to link the binary covariates to the DPM model, $X_{ij}=I(X_{ij}^\star>0)$ for $j=1,2,4,5,6$.
For identification, we restrict the variance of $X_{i1}^\star$, $X_{i2}^\star$, $X_{i4}^\star$, $X_{i5}^\star$, $X_{i6}^\star$ to be $1$. Also, with responses taking values only in $[0,1]$, we use the Tobit model \citep{tobin1958estimation} to link the response to a continuous latent variable $Y^\star$ whose support is $(-\infty,\infty)$, and the observed variable is related to the continuous variable as $\tilde{Y}_{it}=Y^\star_{it}$~if $0\leq Y^\star_{it}\leq 1$, $\tilde{Y}_{it}=0$ if $Y^\star_{it}<0$ and, $\tilde{Y}_{it}=1$ if $Y^\star_{it}>1$.

Then $ (\bX_i^{\star T},Y^\star_{i0})^T\sim \mbox{Normal}(\btheta_{i0},\bSigma_0)$ and $Y_{it}^\star\sim \mbox{Normal}(\calZ_{it}^T\btheta_{i2},\sigma_2^2)$.

After fitting the model to the HP data, diagnostic checks revealed evidence against the normality assumption. Therefore, we use scale mixtures of normals to accommodate the heavier-tailed residual distributions, such that
\beqn\label{scale_mixture}
\log(\delta_{it})|\Theta_i,\bX_i,Y_{it-1},A_{it} &\sim& \mbox{Normal}(\calX_{it}^T\btheta_{i1},\lambda_{it1}\sigma_1^2)\nonumber\\
Y_{it}^\star|\Theta_i,\bX_i,Y_{it-1},\delta_{it} &\sim& \mbox{Normal}(\calZ_{it}^T\btheta_{i2},\lambda_{it2}\sigma_2^2)\nonumber
\eeqn
where $\lambda_{it1}\sim\text{Inv-Gamma}(\nu_1/2,\nu_1/2)$, $\lambda_{it2}\sim\text{Inv-Gamma}(\nu_2/2,\nu_2/2)$. Also, as shown in Figure \ref{f:subject}, there are number of observations with $y_{it}=y_{it-1}$. To account for this, we adjust the disease-progression model such that $Y_{it}$ can have excess probability $p_{it}$ on $y_{it-1}$,
$$f(y|Y_{it-1}=y_{it-1})=p_{it}\mathbbm{1}_{y_{it-1}}(y)+(1-p_{it})\phi^\star(y|\calZ_{it}^T\btheta_{i2},\lambda_{it2}\sigma_2^2)$$
where $p_{it}=\Phi(\calZ_{it}^T\btheta_{i3})$, $\mathbbm{1}_{y_{it-1}}(\cdot)$ is the indicator function with a point mass at $y_{it-1}$, and $\phi^\star(y|\calZ_{it}^T\btheta_{i2},\lambda_{it2}\sigma_2^2)$ is the density of the response variable $\tilde{Y_{it}}$ in the Tobit model with mean $\calZ_{it}^T\btheta_{i2}$ and variance $\lambda_{it2}\sigma_2^2$, which corresponds to a normal distribution for the latent response variable $Y^\star_{it}$.
We assign the Dirichlet process prior for the distribution of $\Theta_i = \{\btheta_{i0},\btheta_{i1},\btheta_{i2}, \btheta_{i3}\}$. The hyperparameter values and MCMC details are described in the supplemental material. The supplemental materials also include model comparisons and goodness of fit diagnostics. We find that the DPM model described in this section with $L=10$ mixture components fits well and outperforms simpler methods. Therefore we use this model for the remainder of the analysis.

\subsection{Summarizing the fitted model}\label{s:comparisons}

The posterior mean and standard deviation for average of $(\btheta_{1l}^{\star T},\btheta^{\star T}_{2l},\btheta_{3l}^{\star T})^T$ weighted by the mixture probabilities $w_l$ for $l=1,2,\ldots,10$ are listed in Table \ref{t:fit}. As expected, the recommended recall interval ($A_{it}$) is the most important factor to determine the actual recall time ($\delta_{it}$), and current disease status ($Y_{it-1}$) is the most important predictor to predict the disease status during next visit ($Y_{it}$). The disease progression between two visits is associated with the actual time between two visits, and the significantly positive interaction effect between current disease status and actual recall time on $Y_{it}^\star$ indicates that time effect is larger for subjects with worse disease status. Also, most of the baseline covariates have significant effect on either the actual recall time or disease progression.

\begin{table}[ht]
\centering
\caption{\small{The posterior mean $\times100$ and standard deviation $\times100$ for weighted average of random effects with $\log(\delta_{it})$, $Y_{it}^\star$ and $\Phi^{-1}(p_{it})$ as responses, respectively. The posterior mean with ``$\star$'' represents the corresponding $95\%$ credible intervals that excludes zero.}}
\label{t:fit}
\begin{tabular}{lcclcccc}
\hline
		Response& \multicolumn{2}{c}{Recall time ($\log(\delta_{it})$)} & & \multicolumn{2}{c}{PMU ($Y_{it}^\star$)} & \multicolumn{2}{c}{Prob equal ($\Phi^{-1}(p_{it})$)}\\
		& Mean & SD & & Mean & SD & Mean & SD \\
		\hline
		Intercept & 84.63$^\star$ & 6.41 & Intercept & -1.76$^\star$ & 0.38 & -47.83$^\star$ & 9.10 \\
		Gender & -0.41 & 1.43 & Gender & -0.16 & 0.11 & 6.84 & 3.44 \\
		Race & 24.75$^\star$ & 2.81 & Race & -0.25 & 0.20 & 1.42 & 5.33 \\
		Age & -10.41$^\star$ & 0.92 & Age & 0.47$^\star$ & 0.08 & -1.08 & 2.55 \\
		Diabetes & -4.16 & 4.98 & Diabetes & -0.36 & 0.26 & -13.87 & 6.99 \\
		Smoking & -17.87$^\star$ & 3.02 & Smoking & 0.98$^\star$ & 0.23 & -9.89 & 6.77 \\
		Insurance & -23.10$^\star$ & 5.65 & Insurance &  2.44$^\star$ & 0.30 & 7.71 & 7.66 \\
		$Y_{it-1}$ & -37.24$^\star$ & 8.43 & $Y_{it-1}$ & 89.67$^\star$ & 0.72 & 92.70$^\star$ & 16.68 \\
		$\log(A_{it})$ & 63.92$^\star$ & 3.45 & $\log(\delta_{it})$ & 1.33$^\star$ & 0.17 & 16.36$^\star$ & 4.37 \\
		Gender$*\log(A_{it})$ & 0.21 & 0.67 & Gender$*\log(\delta_{it})$ & 0.15$^\star$ & 0.05 & 3.36$^\star$ & 1.54 \\
		Race$*\log(A_{it})$ & -11.95$^\star$ & 1.32 & Race$*\log(\delta_{it})$ & 0.00 & 0.09 & -3.72 & 2.49 \\
		Age$*\log(A_{it})$ & 3.12$^\star$ & 0.45 &	Age$*\log(\delta_{it})$ & -0.03 & 0.04 & 0.07 & 1.25 \\
		Diabetes$*\log(A_{it})$ & 2.33 & 2.73 & Diabetes$*\log(\delta_{it})$ & 0.26$^\star$ & 0.12 & 7.62$^\star$ & 3.04 \\
		Smoking$*\log(A_{it})$ & 9.47$^\star$ & 1.63 & Smoking$*\log(\delta_{it})$ & -0.14 & 0.11 & 10.50$^\star$ & 3.33 \\
		Insurance$*\log(A_{it})$ & 10.63 & 3.14 & Insurance$*\log(\delta_{it})$ & -1.28$^\star$ & 0.15 & 1.31 & 3.72 \\
		$Y_{it-1}*\log(A_{it})$ & 17.76$^\star$ & 4.05 & $Y_{it-1}*\log(\delta_{it})$ & 3.61$^\star$ & 0.31 & 2.77 & 7.52 \\
		\hline
\end{tabular}
\end{table}

\subsection{Summarizing the fitted policy}\label{s:results}
While analyzing HP data, we use the policy in (\ref{linearpolicy}) with $a_1=3$ and $a_2=9$, and a linear combination of $q=4$ features representing
standardized age ($X_{3}$), diabetes status ($X_{4}=0$ for subjects without diabetes, and $X_{4}=1$ for subjects with diabetes), non-compliance (via $\log(|\delta_{t-1}-A_{t-1}|+1)$) and disease status (via $Y_{t-1}$). As the scale of $Y_{t-1}$ is much smaller than the other three features, we use $10Y_{t-1}$ in the risk score to get more stable estimates of the feature weights:
$$R_{t}=X_{3}\alpha_1+X_{4}\alpha_2+\log(|\delta_{t-1}-A_{t-1}|+1)\alpha_3 + 10Y_{t-1}\alpha_4.$$
We have also tried replacing diabetes status ($X_{4}$) with gender ($X_{1}$) or smoking status ($X_{5}$) in the construction of the risk score, and this does not improve the value $V$. We define the utility function as
the reduction in proportion of unhealthy sites in 5 years from baseline  $U(\bH) = Y_{0}-Y_{T_{60}}$, where $Y_{T_{60}}$ is the response for a subject in 5 years (60 months) since baseline visit (if no visit occurs at the exact time point, interpolation is used to estimate the response value), and control the cost by constraining average recommended recall time to be $C(\balpha)=6$ months. We use $5,000$ iterations in Gibbs sampling, and discard first $3,000$ burn-in samples to obtain $2,000$ posterior samples by fitting the DPM model. We randomly draw 100 posterior samples of $g$ and $\calS$ and estimate optimal policy in terms of $\balpha_{opt}$ given each selected posterior draw.

\begin{figure}
\caption{\small{Posterior distribution (5th, 25th, 50th, 75th, 95th percentiles) of optimal feature weights $\balpha_{opt}$ for the HP data analysis.  The features are age (standardized), diabetes (with diabetes=1 and without diabetes=0) non-compliance ($\log(|\delta_{t-1}-A_{t-1}|+1))$, and current response ($Y_{t-1}$).}}\label{f:alpha_HP}
\centering
\includegraphics[height=0.6\textwidth]{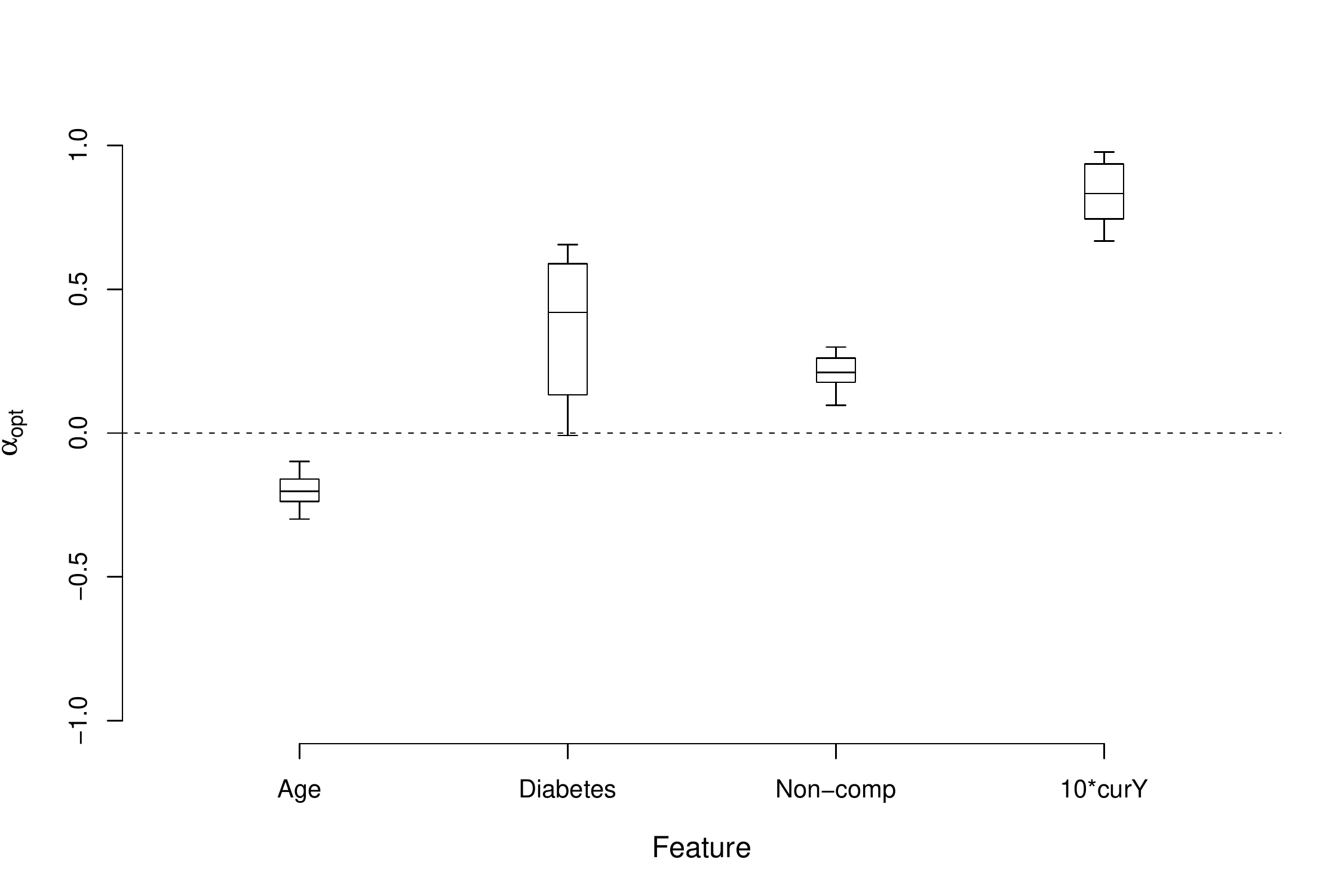}
\end{figure}

Figure \ref{f:alpha_HP} plots the posterior of $\balpha_{opt}$. The posterior mean weights for diabetes and non-compliance are positive. This suggests that subjects with diabetes and non-compliers should be recommended to come back to the dental clinic in a shorter time, reconfirming earlier findings on the link between diabetes and PD \citep{mealey2006diabetes}, and between recall compliance and medium to long-term PD therapy \citep{fenol2010compliance}. However, the posterior distribution of the weight for diabetes has a large variance. The weight for the current disease status is significantly positive with high value, which indicates that the current disease status is an important feature in determining recommendation decision, and the subject with a higher proportion of diseased sites has a higher risk score and should be assigned shorter recall time. The negative estimated weight for age indicates that younger subjects should have more frequent visits. It may be that younger subjects are less stable, and so providing more visit opportunities to younger subjects might improve population-level benefits.

We also estimate one final optimal policy $\balpha_{opt}$ averaging over uncertainty in $g$ and $\calS$, which gives the risk score function $R_{t}=-0.17X_{3}+0.50X_{4}+0.22\log(|\delta_{t-1}-A_{t-1}|+1) + 8.2Y_{t-1}$ with a threshold $1.06$. This risk score function suggests that subjects with diabetes tend to have higher risk than subjects without diabetes, and the disease status is the most important feature that decides the recommended recall time. For example, a subject with diabetes, average age ($X_{3}=0$) and perfect compliance ($\delta_{t-1}=A_{t-1}$) should come back in 3 months if his/her proportion of unhealthy sites is higher than $6.8\%$. A subject without diabetes and with average age and perfect compliance should come back in 3 months if his/her proportion of unhealthy sites is higher than $12.9\%$.

The value corresponding to the estimated optimal policy is $V(\balpha_{opt})=-0.0102$ with standard error $0.0002$, which is estimated by Monte Carlo Simulation with $100,000$ simulated subjects. Compared to the `baseline' policy which recommends $A_{t}=6$ months between visits for all subjects and $t$ and with estimated $V=-0.0170$ with standard error $0.00018$, the utility value averaging over the entire distribution of $\bH$ increases by about $40\%$. This is a substantial improvement, especially when the improvement of expected value is multiplied by the number of people in the population. 


Furthermore, to explore the effects of choosing a linear function of the subject characteristics and current disease status as the risk score, we updated the policy with a quadratic term of current disease status (the most important feature) in constructing the risk score. The estimated risk score is $R_{t}=-0.38X_{3}+0.55X_{4}+0.25\log(|\delta_{t-1}-A_{t-1}|+1) + 3.2Y_{t-1}+63Y_{t-1}^2$, with the value $-0.0101$, which is very close to the value $-0.0102$ corresponding to the optimal policy under the class of policies with only linear function of the features. Hence, we advocate using only linear features for this dataset.

\section{Conclusions}\label{s:con}
Motivated to address the shortcomings of the classic 6-month recall interval in periodontal treatment allocations, we present a policy-optimized recommendation engine using BNP that exhibits superior performance, compared to alternatives. We show using simulation studies that the proposed method provides a valid posterior inference, and can reliably identify the optimal policy. Applying the method to the HP data, we find that the optimal policy recommends more frequent visits for young, unhealthy non-compliers, and that following this policy could lead to a substantial reduction in PD.

A number of computerized periodontal risk assessment tools are currently available \citep[][]{page2003longitudinal, persson2003assessing}. For example, the Cigna PD self-assessment tool available at \url{https://www.cigna.com/healthwellness/tools/periodontal-quiz-en} considers subject-specific inputs through a questionnaire, and combines information from the PD fact sheet of the \textit{American Academy of Periodontology} to calculate a simple ordinal risk score (low, low to moderate, moderate, or high), without any guidance towards recall intervals. There exists a number of popular chairside software in practice-based dentistry (such as Patterson's Eaglesoft\textregistered) that record and display data. Supplementing these tools with an evidence-based recommendation system for periodontal recalls would aid practitioners.

A limitation of our analysis is that we use periodontal pocket depth (PPD) rather than the most reliable endpoint \citep{AAPclassify}, the clinical attachment level (CAL). Site-level full mouth CAL assessment in a practice-based observational data setting like ours is time-consuming and technically demanding \citep{michalowicz2013change}. For example, in the HP dataset, CAL is computed only for the mid-buccal and mid-lingual sites, whereas, the PPD is calculated for all 6 sites on each tooth (if that tooth is present). Also, since CAL is computed from two other measures, it is more prone to error, and less reproducible than PPD \citep{osbom1992comparison, hill2006study}. Hence, we considered thresholded site-level PPD in addition to missing tooth to compute the proportion subject-level endpoints. The missing tooth in our analysis is assumed missing due to past incidence of PD, and the error generated from the apparent misclassification of the missingness source (such as, tooth falling out due to mechanical injury) is usually negligible while analyzing large observational databases. Should CAL and PPD measures become available for all sites (in other databases), our framework can readily incorporate this information. In addition, to reduce computational burden, our present policy only considers recall intervals of 3 and 9 months. Our method can be easily extended to more than two possible actions by adding thresholds to the risk score. Computationally, estimating an optimal threshold parameter should be similar to estimating a feature weight. 
Therefore, the proposed decision framework is quite general, and can be adapted to the specific needs of the practitioner.

To the best of our knowledge, this is the \textit{first study} to cast the century-old debate on periodontal recall intervals into a DTR stochastic framework. Our recommendation tool is derived from a specific US midwestern population, and it's generalizability should be tried with caution. Longitudinal PD databases from other practice-based settings (such as Kaiser Permanente\textregistered) maybe combined with the current HP database to refine findings. Furthermore, our present recall engine is geared exclusively towards PD assessment; and do not include (dental) caries risk, although evidence suggest that they may occur simultaneously \citep{mattila2010prevalence}. These are potential directions for future work, to be pursued elsewhere.



\begin{singlespace}
\bibliographystyle{biom}
\bibliography{policy}
\end{singlespace}

\end{document}